\documentclass[a4paper,11pt]{article}
\usepackage{jhepmod} 
\usepackage{lineno}

\allowdisplaybreaks[4]

\arxivnumber{2412.20474} 

\title{Cosmological stimulated emission}

\author{Atsuhisa Ota}
\affiliation{Department of Physics and Chongqing Key Laboratory for Strongly Coupled Physics, \\
Chongqing University, Chongqing 401331, People's Republic of China}

\emailAdd{aota@cqu.edu.cn}

\abstract{We study stimulated emission and absorption of gravitons in a squeezed vacuum state immersed in a thermal radiation bath. Employing one‐loop interaction‐picture perturbation theory, we track the time evolution of the graviton number operator and its expectation value in the squeezed vacuum, which characterizes the inflationary graviton state. In a Minkowski background with a thermal bath as a toy example, we demonstrate that the net graviton emission or absorption rate depends sensitively on the initial squeezing parameters. As a thought experiment, we consider LIGO/Virgo-like detectors operating in radiation at temperatures of order 0.1\,GeV and find that graviton occupation numbers at frequencies of order 100\,Hz can be significantly enhanced, suggesting a novel mechanism for amplifying gravitational-wave signals. Although these conditions exceed current experimental capabilities, they point toward potential future advances in detection. Extending our analysis to an expanding, radiation-dominated universe, we show that subhorizon gravitons undergo stimulated absorption, while superhorizon modes exhibit secular logarithmic growth, indicating the breakdown of perturbative methods and motivating further investigation. These findings open a new direction for exploring graviton coherence effects in realistic cosmological and laboratory settings.

}

\begin{document}
\maketitle

\section{Introduction}
\label{sec:intro}

Stimulated emission enhances boson production in a mode that is already occupied by identical quanta. In laser physics, for example, an incoming photon of frequency~$\omega$ interacts with an excited atomic electron, inducing a transition to a lower energy level and emitting a second photon with the same frequency, direction, and polarization~\cite{Weinberg:2015QM}. This process underlies coherent beam amplification.

In this paper, we investigate whether an analogous mechanism can operate for gravitons in a cosmological medium. Our primary motivation is to study the secondary evolution of primordial gravitons from their generation during inflation to their re‑entry and observational signatures today. It is well established that graviton fluctuations generated quantum mechanically during inflation exist outside the horizon as squeezed vacuum states~\cite{Grishchuk:1990bj}, remain frozen on superhorizon scales, and later re‑enter the horizon to contribute, for instance, to the B‑mode polarization of the cosmic microwave background~\cite{Mukhanov:2005sc}.

Recently, the author and collaborators studied graviton dynamics in a dense thermal plasma by minimally coupling gravitons to a massless scalar field as a proxy for radiation~\cite{Ota:2023iyh}. Although the graviton scattering cross section is suppressed by the Planck scale, the large particle density of the plasma can yield a non‑negligible net effect. Prior work suggested a secular (i.e., cumulative) growth of infrared graviton modes via repeated plasma interactions, a result that has been independently confirmed in alternative setups~\cite{Frob:2025sfq}. However, such growth may reflect unphysical gauge artifacts or the breakdown of perturbation theory. Here, we revisit this phenomenon and isolate the genuine physical contribution.

Our goal in this paper is not to resolve the debate over infrared secular growth itself, but rather to isolate an unambiguously gauge-invariant sector—namely sub-horizon graviton modes (or, equivalently, modes in Minkowski spacetime)—and to demonstrate that their loop-induced evolution admits a transparent analogy to stimulated emission in quantum optics.  The infrared issue is therefore deferred to future work.

To capture the relevant quantum coherence, we compute the time evolution of the graviton number operator (or energy density) in the in–in (Schwinger–Keldysh) formalism, rather than focusing on the strain power spectrum. This framework distinguishes spontaneous graviton production and subsequent kinematical scattering, as described by Boltzmann kinetic theory~\cite{Ghiglieri:2015nfa,Ghiglieri:2020mhm}, from coherent, stimulated emission driven by initial squeezed states.

The novel contributions of this work are:
\begin{itemize}
  \item A quantitative demonstration of stimulated graviton emission in a thermal medium arising from an initial squeezed vacuum state.
  \item A clear separation between spontaneous and stimulated graviton production, highlighting quantum coherence effects beyond the mixed‑state approximation.
  \item Perturbative evaluation of the stimulated emission across relevant parameter regimes, such as Minkowski spacetime and sub‑horizon scales at reheating.
\end{itemize}

These findings open a new channel for graviton amplification in the early universe and furnish a robust framework for evaluating its observational consequences during the radiation-dominated era. A systematic treatment of infrared secular terms, whether via symmetry-based arguments or hard-thermal-loop resummation, will be presented elsewhere.

The rest of this paper is organized as follows. In Section~\ref{sec:stimfdthory} we review the general field‑theoretic formulation of stimulated emission in a simple QED example. Section~\ref{sec:setup} presents our cosmological setup, introducing the free graviton theory, the interaction Hamiltonian, the time‑dependent number operator, and the squeezed vacuum initial state. In Section~\ref{sec:emission} we derive the 1‑loop expression for the stimulated graviton emission rate and distinguish the spontaneous and coherent contributions. Section~\ref{sec:minkowski} applies this formalism to a Minkowski background to build physical intuition, while Section~\ref{sec:rad‐dom} extends the analysis to a radiation‑dominated universe and discusses both subhorizon and superhorizon limits. We conclude in Section~\ref{sec:conclusion} with a summary of our main results and an outlook. Technical details and extended derivations are collected in Appendices~A–C.

\section{Stimulated emission in field theory}\label{sec:stimfdthory}
Stimulated emission is often discussed in the context of non-relativistic quantum mechanics. Since we consider this effect in a cosmological setting, we first examine it from a field-theoretical perspective (see also Ref.~\cite{Aleksandrov:2022rgg}).

As an illustrative example, consider a photon field in Minkowski spacetime, $\hat{A} = u \hat{d} + u^* \hat{d}^\dagger$, with the positive-frequency mode function $u$ and the annihilation operator $\hat{d}$. For notational simplicity, we suppress the polarization index, Lorentz index, and momentum, and we write $[\hat d, \hat d^\dagger]=1$. A rigorous treatment is postponed for gravitons. The photon number operator is defined as $\hat N \equiv \hat d^\dagger \hat d$, which is well-defined only in flat spacetime. We study the evolution of $\hat N$ under photon–electron interactions in quantum electrodynamics (QED). The interaction Hamiltonian is given by $\hat H_I = -e \hat j \hat A$, where $\hat j$ is the U(1) current and $e$ is the coupling constant.

The time evolution of the number operator in the interaction picture is given by~\cite{Weinberg:2005vy}
\begin{align}
    \sum_{n=0}^\infty \hat N_n(\tau) &=\sum_{n=0}^{\infty} i^n \int^\tau d\tau_1 \cdots \int^{\tau_{n-2}} d\tau_{n-1} \int^{\tau_{n-1}} d\tau_n \notag \\
    &\quad \times \left[\hat H_I(\tau_n), \left[\hat H_I(\tau_{n-1}), \cdots \left[\hat H_I(\tau_1), \hat N(\tau)\right]\cdots \right]\right], \label{hintcom}
\end{align}
where $\hat N$ is in the interaction picture.

We evaluate the ensemble average of Eq.~\eqref{hintcom} for a given quantum state $\hat \varrho$ of the photon–electron system. For simplicity, we assume that the photon and electron states are separable: $\hat \varrho = \hat \varrho_{\gamma} \otimes \hat \varrho_{\mathrm{e}}$. For stimulated emission, $\hat \varrho_{\rm e}$ exhibits population inversion, and $\hat \varrho_\gamma$ is chosen such that the photon frequency matches the atomic transition energy. Thus, when the quantum states are \textit{fine‑tuned}, stimulated emission or absorption occurs. The laser mechanism is just one example; bound-state quantum mechanics is not essential.

Now, let us explicitly evaluate the change in photon number induced by $\hat H_I$. Neglecting momentum, polarization, and spacetime indices, the algebraic calculation is straightforward. First, the commutation relation between the interaction Hamiltonian and the number operator is
\begin{align}
    \left[\hat H_I(\tau_1), \hat N\right]
    &= \left[-e \hat A_1 \hat j_1, \hat d^\dagger \hat d\right]
    = -e \left[u_1 \hat d + u_1^* \hat d^\dagger, \hat d^\dagger \hat d\right] \hat j_1 \notag \\
    &= -e \left(u_1 \hat d - u_1^* \hat d^\dagger\right) \hat j_1,
\end{align}
where we have used only the commutation relation $[\hat d, \hat d^\dagger] = 1$. Similarly, an additional commutator gives
\begin{align}
    \left[\hat H_I(\tau_2), \left[\hat H_I(\tau_1), \hat N\right]\right]
    &= e^2 \left[\left(u_2 \hat d + u_2^* \hat d^\dagger\right)\hat j_2, \left(u_1 \hat d - u_1^* \hat d^\dagger\right)\hat j_1\right].
\end{align}
Note that for $[A_i, B_j] = 0$, one has
\begin{align}
    [A_2 B_2, A_1 B_1]
    &= \tfrac{1}{2}(A_1 A_2 + A_2 A_1)\,[B_2, B_1]
    + [A_2, A_1]\,\tfrac{1}{2}(B_1 B_2 + B_2 B_1).
\end{align}
Hence we find
\begin{align}
    \left[\hat H_I(\tau_2), \left[\hat H_I(\tau_1), \hat N(\tau)\right]\right]
    &= -e^2\,(u_2 u_1^* + u_2^* u_1)
       \tfrac{\hat j_1 \hat j_2 + \hat j_2 \hat j_1}{2} \notag \\
    &\quad - e^2
       \Bigl(u_1 u_2\,\hat d \hat d
       - u_1^* u_2^*\,\hat d^\dagger \hat d^\dagger
       + (u_1 u_2^* - u_1^* u_2)\,\tfrac{\hat d^\dagger \hat d + \hat d \hat d^\dagger}{2}\Bigr)
       \bigl[\hat j_1, \hat j_2\bigr]. \label{double][][}
\end{align}

The expectation value for $n=1$ vanishes because $\hat N_1$ is linear in $(\hat d,\hat d^\dagger)$. 
Eq.~\eqref{double][][} yields the next-to-leading-order correction: $  \hat N_2 =   \hat N_{{\rm spon}} +   \hat N_{{\rm stim}}$, with
\begin{align}
      \hat N_{{\rm spon}} &= e^2 \int^\tau d\tau_1  \int^{\tau_1} d\tau_2 \, \frac{\hat j(\tau_1) \hat j(\tau_2)+\hat j(\tau_2) \hat j(\tau_1)}{2}  \notag 
    \\
    &\quad \times (u(\tau_2) u(\tau_1)^* + u(\tau_1) u(\tau_2)^*), \label{dN2spon} \\
      \hat N_{{\rm stim}} &= e^2 \int^\tau d\tau_1  \int^{\tau_1} d\tau_2 \, \left[ \hat j(\tau_1),\hat j(\tau_2) \right] \notag 
    \\
    &\quad \times \big(   ( u(\tau_1) u(\tau_2)^*    - u(\tau_1)^* u(\tau_2) )\frac{\hat d \hat d^\dagger+  \hat d^\dagger \hat d}{2}  \notag 
    \\
    &\qquad  + u(\tau_1)u(\tau_2) \hat d^2 - u(\tau_1)^* u(\tau_2)^* \hat d^{\dagger 2} \big), \label{dN2stim}
\end{align}
The photon-number evolution depends solely on the electron state, so Eq.~\eqref{dN2spon} represents spontaneous emission, whereas Eq.~\eqref{dN2stim} depends on the initial photon state.  Consider a separable photon–electron state, $\hat \varrho = \hat \varrho_\gamma \otimes \hat \varrho_{\rm e}$, with the $N$-photon state $|N \rangle \equiv \frac{(\hat{d}^\dagger)^N}{\sqrt{N!}} |0 \rangle$ satisfying $\hat d |0\rangle =0$, and $\hat \varrho_\gamma = |N \rangle \langle N|$.  In this number eigenstate, the last term vanishes, and Eq.~\eqref{dN2stim} yields 
\begin{align}
    {\rm Tr}\left[\hat  \varrho   \hat N_{{\rm stim}} \right]  =& e^2\int^\tau d\tau_1 \int^{\tau_{1}} d\tau_{2} (u_1 u_2^* - u_1^* u_2)N     {\rm Tr}\left[ \hat \varrho_{\rm e} \left[  \hat j_1 ,  \hat j_2 \right]\right],\label{stimN}
\end{align}
which describes stimulated emission or absorption.  More precisely, Eq.~\eqref{dN2stim} contains a vacuum-stimulated term proportional to $\hat d \hat d^\dagger$.

In contrast to number eigenstates, squeezed vacua exhibit a distinct feature.  A squeeze operator $\hat S(z)$ is defined by 
\begin{align}
    \hat S(z) \equiv \exp\left( \frac{z^* \hat d^2 - z (\hat d^\dagger)^2}{2}\right),~z = r e^{i\theta}.
\end{align}
It generates the squeezed vacuum state $|\psi\rangle = \hat S(z) |0\rangle$.  The annihilation operator $\hat d_{\psi}$ satisfying $\hat d_\psi |\psi\rangle=0$ is related to $\hat d$ by
\begin{align}
 	0= \hat S(z)^\dagger \hat d_\psi |\psi\rangle = \hat S(z)^\dagger \hat d_{\psi} \hat S(z) |0\rangle\to   \hat d   = \hat S(z)^\dagger \hat d_\psi \hat S(z),
\end{align}
which gives, $\hat d_\psi  = \cosh r \hat d   - e^{i\theta}\sinh r \hat d^\dagger$.
This is recast into
\begin{align}
    \hat d = \mu \hat d_{\psi}   + \nu \hat d^\dagger_{\psi },
\end{align}
with $\mu = \cosh r$, $\nu=e^{i\theta}\sinh r$, and $ |\mu|^2 - |\nu|^2 =1$.  Hence
\begin{align}
    \langle \psi |\hat N |\psi \rangle  = |\nu|^2.\label{ppnu}
\end{align} 
Thus, the squeezed vacuum acts as an excited state with occupation number $|\nu|^2$.  Pair correlations are nonzero:
\begin{align}
    \langle \psi | \hat d \hat d |\psi \rangle &=\langle \psi | (\mu \hat d_\psi + \nu \hat d_\psi^\dagger) (\mu \hat d_\psi + \nu \hat d_\psi^\dagger)|\psi \rangle = \mu\nu,
    \\
    \langle \psi | \hat d^\dagger \hat d^\dagger |\psi \rangle &=\langle \psi | (\mu^* \hat d_\psi^\dagger + \nu^* \hat d_\psi ) (\mu^* \hat d_\psi^\dagger + \nu^* \hat d_\psi)|\psi \rangle = \nu^* \mu^*.
\end{align}
These non-vanishing pair correlations add extra terms to Eq.~\eqref{stimN}:
\begin{align}
    {\rm Tr}\left[\hat  \varrho   \hat N_{2, {\rm stim}} \right]  \simeq e^2 \int^\tau d\tau_1 \int^{\tau_{1}} d\tau_{2}  \left( (u_1 u_2^* - u_1^* u_2)|\nu|^2 -\mu^*\nu^* u_1^* u_2^*+ \mu \nu u_1 u_2       \right)  {\rm Tr} \left[ \hat \rho_{\rm e} \left[  \hat j_1 ,  \hat j_2 \right]\right].
\end{align}
Thus, squeezed-vacuum correlations further enhance stimulated emission or absorption, indicating that Bose enhancement extends beyond identical number eigenstates.

\medskip

In cosmology, gravitational-wave production is often used to constrain early-universe models (see Ref.~\cite{Caprini:2018mtu} and references therein), corresponding to spontaneous emission in Eq.~\eqref{dN2spon}.  In laser media, greater spontaneous emission generally implies stronger stimulated emission via the relation between absorption and emission coefficients~\cite{Weinberg:2015QM}.  Analogously, mechanisms that produce substantial gravitational waves suggest potential for significant cosmological stimulated emission of gravitons.  Note that squeezed vacua are crucial here, since inflationary gravitons exist in such a state~\cite{Grishchuk:1989ss, Albrecht:1992kf}.

\section{Setup}
\label{sec:setup}

\begin{figure}
\centering
	\includegraphics[width = 0.4\linewidth]{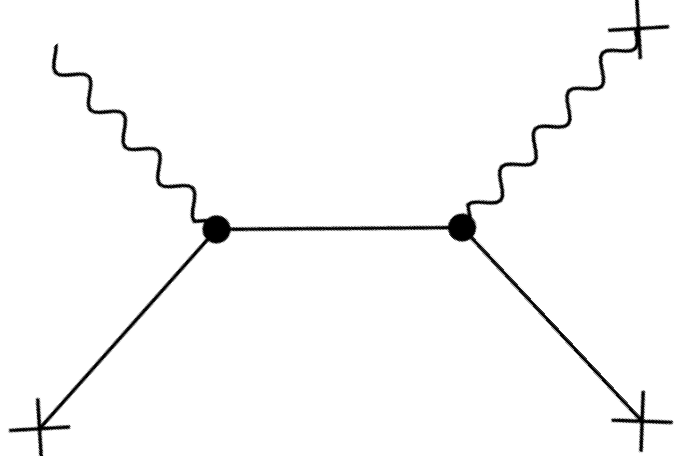}
	\caption{A diagrammatic representation of stimulated emission for gravitons. Solid and wavy curves represent the scalar field $\chi$ and graviton mode $h_{ij}$, respectively. The gravitational coupling constant is $M_{\rm pl}^{-1}$. A cross on an external leg implies the thermal bath for $\chi$ and the excited state for gravitons. The internal line represents the retarded Green function for $\chi$. This figure illustrates the linear response of the thermal bath to excited gravitons. The diagram is suppressed by $M_{\rm pl}^{-2}$, but the thermal bath effect is enhanced by the radiative pressure $P_\chi \sim T^4$.}
	\label{fig0}
\end{figure}

In the previous section, we provided an illustrative QED example to formulate stimulated emission in field theory and discussed its distinctive feature in the squeezed vacuum state.  
In this section, we describe our setup for gravitons in a cosmological setting.  
We draw an analogy between photons and gravitons and introduce the key quantities.
See Fig.~\ref{fig0} for a schematic picture of the stimulated emission in our mind.

As an example of cosmological stimulated emission, consider gravitons minimally coupled to a massless scalar field $\chi$ in a spatially flat Friedmann–Lemaître–Robertson–Walker (FLRW) background.  
The field $\chi$ is free but is initially in a local thermal state described by a canonical ensemble with comoving inverse temperature~$\beta$.  
Because a canonical ensemble lacks population inversion---that is, an excited-state surplus---we do not anticipate stimulated emission in the usual sense in this setup.
However, we reveal that the squeezed state of gravitons plays an alternative role: both stimulated emission and absorption can occur.  
We define all Hamiltonians with respect to conformal time, so $\beta$ is understood as a comoving inverse temperature. 

We consider a free scalar field as a proxy for the thermal radiation bath and therefore neglect self-interactions inside the medium.  
In a weakly-coupled plasma with coupling constant $g$, interactions generate a mean–free–path scale $\ell_{\rm mfp}$ obeying $\ell_{\rm mfp}/\beta \sim (g^{4}\ln (1/g) )^{-1} \gg 1$~\cite{Arnold:2003zc}.  
Hence the thermal scale $\beta$ remains the fastest dynamical time-scale in the system.  
As we shall see below, the stimulated emission occurs within a short interval $\Delta\tau \sim \beta$, so the free-field approximation is self-consistent.
In other words, non-hydrodynamical regime is considered in the present setup as the free scalar field does not form a fluid.
By contrast, in the strong-coupling regime $g \sim \mathcal{O}(1)$, holographic calculations give $\ell_{\rm mfp}/\beta = \mathcal{O}(1)$~\cite{Kovtun:2004de}; plasma effects may then compete with the temperature scale and would require a more detailed treatment.  
In what follows we therefore restrict ourselves to the weak-coupling case or more extreme scenarios of dark radiation.
Such a situation is not artificial at all as it applies to cosmological neutrinos after neutrino decoupling.
Additional contributions from the hydrodynamical regime will be discussed based on the Kubo formula in future work.

\subsection{Graviton free theory}
A graviton $h_{ij}$ is defined as the quantized traceless, transverse perturbation around a flat FLRW metric:  
Consider an FLRW background spacetime and traceless, transverse perturbations:
\begin{align}
	ds^2 &= a(\tau)^2 \left(-d\tau^2 + \gamma_{ij} \,dx^i dx^j \right), \\
	\gamma_{ij} &\equiv \delta_{ij} + \frac{2}{M_{\rm pl}}\,h_{ij} + \frac{2}{M_{\rm pl}^2}\,h_{ik}\,h^k{}_j + \cdots, \\
	h^i{}_i &= 0, 
	\quad
	\partial_i h^i{}_j = 0.
	\label{eq:const}
\end{align}
By expanding the Einstein–Hilbert action to second order in $h^i{}_j$, one finds
\begin{align}
	S_h &= \int d\tau\,L_h[h^i{}_j,\,h'^i{}_j,\,\tau], \\
	L_h[h^i{}_j,\,h'^i{}_j,\,\tau] &\equiv  \frac{1}{2} \int d^3x\,a(\tau)^2 \Bigl((h'^i{}_j)^2 - (\partial_k h^i{}_j)^2\Bigr).
	\label{lagdef}
\end{align}
We define the conjugate momentum
\begin{align}
	\pi^j{}_i \equiv \frac{\delta L_h[h^i{}_j,\,h'^i{}_j,\,\tau]}{\delta h^i{}_j}
	= a^2 \,h'^j{}_i.
\end{align}
One may express the Fourier integrals of $h_{ij}$ and $\chi$ as
\begin{align}
    \hat{h}_{ij}(\tau, \mathbf{x})
    &= \sum_{s=\pm} \int \frac{d^3 k}{(2\pi)^3}\,e^{i \mathbf{k}\cdot\mathbf{x}}\,
       e^{(s)}_{\mathbf{k},ij}\,\hat{h}^{(s)}_{\mathbf{k}}(\tau), \\
    \hat{\chi}(\tau, \mathbf{x})
    &= \int \frac{d^3 k}{(2\pi)^3}\,e^{i \mathbf{k}\cdot\mathbf{x}}\,
       \hat{\chi}_{\mathbf{k}}(\tau),
\end{align}
where the gravitational-wave polarization tensor satisfies 
\begin{align}
	k^i e^{(s)}_{\mathbf{k},ij} &= 0,
	\\
	\delta^{ij} e^{(s)}_{\mathbf{k},ij} &= 0,\\
	e^{(s)}_{\mathbf{k},ij}\,(e^{(s')ij}_{\mathbf{k}})^* &= \delta^{ss'}.
\end{align}
We impose the canonical commutation relation in Fourier space:
\begin{align}
	\bigl[\hat{h}^{(s)}_{\mathbf{k}},\,\hat{\pi}^{(s')}_{\mathbf{k}'}\bigr]
	= i\,\hbar\,\delta_{ss'}\,(2\pi)^3\,\delta(\mathbf{k}+\mathbf{k'}).
\end{align}
Hereafter, we set $\hbar = 1$.  
In the Heisenberg picture, operators evolve while states are fixed at some reference time $\tau_0$.  
We expand the field operators in terms of creation and annihilation operators at $\tau_0$ using mode functions:
\begin{align}
	h^s_{\mathbf{k}}(\tau)
	&= u_k(\tau, \tau_0)\,\hat{d}^s_{\mathbf{k}}(\tau_0)
	+ u_k^*(\tau, \tau_0)\,\hat{d}^{s\dagger}_{-\mathbf{k}}(\tau_0), \\
	\pi^s_{\mathbf{k}}(\tau)
	&= v_k(\tau, \tau_0)\,\hat{d}^s_{\mathbf{k}}(\tau_0)
	+ v_k^*(\tau, \tau_0)\,\hat{d}^{s\dagger}_{-\mathbf{k}}(\tau_0).
\end{align}
The mode functions depend on the choice of initial operators; see Appendix~\ref{sumofmodefunc} for details.

\subsection{Interaction}
Interactions between $\chi$ and $h_{ij}$ originate from the kinetic term of $\chi$:
\begin{align}
	-\frac{1}{2} \int d^4x \, \sqrt{-g} \, g^{\mu\nu} \partial_\mu \chi \, \partial_\nu \chi 
	&\supset   M_{\rm pl}^{-1} \int d^4 x \, a^2 \, h^{ij} \partial_i \chi \, \partial_j \chi,
\end{align}
where $g^{ij} = a^{-2}(\delta^{ij}-h^{ij}+\cdots)$.  Using the Legendre transformation, we obtain the leading-order interaction:
\begin{align}
    \hat{H}_{I} = -\frac{a^2}{M_{\rm pl}} \int d^3 x \, \hat{h}^{ij} \partial_i \hat{\chi} \, \partial_j \hat{\chi}. \label{intHdef}
\end{align}
At one-loop order, the four-point interaction also contributes to the spectrum.  However, by the equivalence principle this one-loop correction cancels, so we ignore it in this paper~\cite{Ota:2023iyh}.  Expanding the interaction Hamiltonian~\eqref{intHdef} in Fourier space gives
\begin{align}
	\hat{H}_I &= -\frac{a^2}{M_{\rm pl}} \int d^3x \, \hat{h}^{ij} \partial_i \hat{\chi} \, \partial_j \hat{\chi} \notag \\
	&= -\frac{a^2}{M_{\rm pl}} \int d^3x \, \int \frac{d^3k \, d^3p_1 \, d^3p_2}{(2\pi)^9} e^{i\mathbf{k} \cdot \mathbf{x} + i\mathbf{p_1} \cdot \mathbf{x} + i\mathbf{p_2} \cdot \mathbf{x}} \hat{h}^{ij}_{\mathbf{k}}(ip_{1i})(ip_{2j}) \hat{\chi}_{\mathbf{p_1}} \hat{\chi}_{\mathbf{p_2}} \notag \\
	&= \frac{a^2}{M_{\rm pl}} \int \frac{d^3k \, d^3p_1 \, d^3p_2}{(2\pi)^9} (2\pi)^3 \delta(\mathbf{k} + \mathbf{p_1} + \mathbf{p_2}) \sum_{s} e^{ij(s)}_{\mathbf{k}} \hat{h}^{(s)}_{\mathbf{k}} p_{1i} p_{2j} \hat{\chi}_{\mathbf{p_1}} \hat{\chi}_{\mathbf{p_2}} \notag \\
	&= -M_{\rm pl}^{-1} \sum_{s} \int \frac{d^3k}{(2\pi)^3} \hat{h}^{(s)}_{\mathbf{k}} \hat{T}^{(s)}_{\mathbf{k}},\label{intHdef3}
\end{align}
where the projected energy-momentum tensor is
\begin{align}
    \hat{T}^{(s)}_{\mathbf{k}} &\equiv -a^2 \int \frac{d^3 l \, d^3 p}{(2\pi)^3} \delta(\mathbf{k} + \mathbf{l} + \mathbf{p}) e^{(s)ij}_{\mathbf{k}} l_i p_j \hat{\chi}_{\mathbf{l}} \hat{\chi}_{\mathbf{p}}. \label{defemts}
\end{align}
Hence, the correspondence is $e \to M_{\rm pl}^{-1}$, $j \to T$, and $A \to h$.

\subsection{Number operator in a time-dependent background}
In an expanding universe without time-translation symmetry, even the free graviton number operator becomes time-dependent.
To set up interaction-picture perturbation theory, we first need to specify the meaning of “free” in this context. In our convention, we refer to gravitons propagating on a fixed cosmological background as “free” if they obey a linearized equation of motion derived from the Einstein-Hilbert action. While such fields are coupled to the time-dependent background metric, we do not regard this coupling as an interaction. This usage is motivated by the structure of perturbation theory: in the interaction picture, “free” fields are those whose Heisenberg equations of motion can be solved exactly, and whose nonlinear self-interactions or couplings to matter are treated perturbatively.

With this definition, we construct the time-dependent graviton number operator by diagonalizing the free Hamiltonian:
\begin{align}
	H_h[h^i{}_j, \pi^j{}_i, \tau] &= \int d^3x \, \pi^j{}_i h'^i{}_j - L_h[h^i{}_j, h'^i{}_j, \tau) \notag \\
	&= \frac{1}{2} \int d^3x \left( \frac{(\pi^i{}_j)^2}{a^2} + a^2 (\partial_k h^i{}_j)^2 \right) \notag \\
	&= \frac{1}{2} \sum_s \int \frac{d^3k}{(2\pi)^3} \left( \frac{\hat{\pi}^{(s)}_{\mathbf{k}} \hat{\pi}^{(s)}_{-\mathbf{k}}}{a^2} + a^2 k^2 \hat{h}^{(s)}_{\mathbf{k}} \hat{h}^{(s)}_{-\mathbf{k}} \right). \label{hamgrav}
\end{align}
The instantaneous annihilation operator~\cite{Kanno:2021vwu} is
\begin{align}
    \hat{d}^{(s)}_{\mathbf k} &\equiv  a \sqrt{\frac{k}{2}} \hat{h}^{(s)}_{\mathbf k} + \frac{i}{a \sqrt{2k}} \hat{\pi}^{(s)}_{\mathbf k},\label{def:instan}
\end{align}
so that
\begin{align}
	H_h[h^i{}_j, \pi^j{}_i, \tau] 
	&= \sum_s \int \frac{d^3k}{(2\pi)^3} k \Bigl( \hat{d}^{(s)\dagger}_{\mathbf{k}} \hat{d}^{(s)}_{\mathbf{k}} + \tfrac{1}{2} [ \hat{d}^{(s)}_{\mathbf{k}}, \hat{d}^{(s)\dagger}_{\mathbf{k}} ] \Bigr), \label{H0def}
\end{align}
where the commutator term is a constant shift.  We then read off the instantaneous number operator:
\begin{align}
	\hat N^{(s)}_{\mathbf k} \equiv \hat d^{(s)\dagger}_{\mathbf{k}} \hat d^{(s)}_{\mathbf{k}}.
\end{align}
Thus, the number operator inherits time dependence through the canonical variables.

\subsection{Quantum states}
The full evolution is specified by the initial quantum state $|\psi\rangle$ and the operator dynamics.  In cosmology, $|\psi\rangle$ is taken to be the vacuum in the distant past during inflation.  Eq.~\eqref{def:instan} defines the instantaneous vacuum $|0\rangle$ at time $\tau$, satisfying $\hat{d}^{(s)}_{\mathbf{k}}|0\rangle = 0$, which minimizes the Hamiltonian~\eqref{H0def}.  Since $|0\rangle$ depends on $\tau$, it differs from the initial state $|\psi\rangle$.  The two are related by a dynamical Bogoliubov transformation,
\begin{align}
    \hat{d}^{(s)}_{\mathbf{k}}  
    = \mu_{k}\,\hat{d}^{(s)}_{\psi,\mathbf{k}} 
    + \nu_{k}\,\hat{d}^{(s)\dagger}_{\psi,-\mathbf{k}}, 
    \label{instdefd}
\end{align}
where the coefficients depend on time through the background evolution.  One then finds
\begin{align}
    V^{-1}\langle \psi|\hat{d}^{(s)\dagger}_{\mathbf{k}} \hat{d}^{(s)}_{\mathbf{k}}|\psi\rangle
    = |\nu_k|^2,
\end{align}
which gives the number of free gravitons produced in the comoving volume $V \equiv (2\pi)^3\delta(0)$ at time $\tau$.

\section{Stimulated graviton emission}
\label{sec:emission}

In the previous section, we introduced the linear theory of gravitons in a general FLRW background and then defined the interaction Hamiltonian and the time-dependent graviton number operator.  
We now evaluate the graviton analogue of Eq.~\eqref{dN2stim}.  
This calculation corresponds to a one-loop perturbative analysis and involves technically intricate details.  
We split the derivation of the final formulas into three steps.

\paragraph{Step 1: Derivation of formal expressions.}

In this first step, we derive the graviton analogue of Eq.~\eqref{dN2stim}.  
Readers who wish to extend this calculation to other fields, such as thermal vectors or fermions, may use the general formulas presented here.  
For notational simplicity, we rewrite the interaction Hamiltonian~\eqref{intHdef3} as
\begin{align}
    \hat{H}_{I} = -M_{\rm pl}^{-1}\,\hat{h}^S\,\hat{T}_S, \label{intHdefsimp}
\end{align}
where $S$ denotes both the polarization index $s$ and the Fourier wavenumber $\mathbf{k}$, with repeated indices summed and integrated.  
Hereafter, we use the notations in Eqs.~\eqref{intHdef3} and \eqref{intHdefsimp} interchangeably.

The leading-order correction ($n=1$ in Eq.~\eqref{hintcom}) is
\begin{align}
    \hat{N}_1^S &= -\,i\,M_{\rm pl}^{-1} \int^\tau d\tau_1 \,\bigl[ \hat{h}^{S_1}(\tau_1), \hat{N}^{S}(\tau) \bigr] \,\hat{T}_{S_1}(\tau_1). \label{spontanous}
\end{align}
The density operator of the graviton–scalar system is separable in a local inertial frame where $h_{ij}$ vanishes.  
The partial trace over $\chi$ can be evaluated there.  Such tadpole diagrams are perturbed in a general frame and cancel the one-loop contribution from the four-point interaction, as noted after Eq.~\eqref{intHdef}.

For $n=2$, the correction splits into stimulated and spontaneous parts:
\begin{align}
   \hat{N}_{2,{\rm stim}}^S(\tau) &=
    -M_{\rm pl}^{-2} \int^\tau d\tau_1 \int^{\tau_1} d\tau_2 \,\bigl[ \hat{T}_{S_2}(\tau_2), \hat{T}_{S_1}(\tau_1) \bigr] \notag\\
    &\quad\times 
     \frac{\hat{h}^{S_2}(\tau_2)\,\bigl[ \hat{h}^{S_1}(\tau_1), \hat{N}^{S}(\tau) \bigr]
          + \bigl[ \hat{h}^{S_1}(\tau_1), \hat{N}^{S}(\tau) \bigr]\,\hat{h}^{S_2}(\tau_2)}{2}, 
    \label{Bogotransf2}\\
    \hat{N}_{2,\rm spon}^S(\tau) &=
    -M_{\rm pl}^{-2} \int^\tau d\tau_1 \int^{\tau_1} d\tau_2  
    \,\frac{\hat{T}_{S_2}(\tau_2)\,\hat{T}_{S_1}(\tau_1) + \hat{T}_{S_1}(\tau_1)\,\hat{T}_{S_2}(\tau_2)}{2}
    \notag\\
    &\quad\times
    \bigl[\hat{h}^{S_2}(\tau_2), \bigl[ \hat{h}^{S_1}(\tau_1), \hat{N}^{S}(\tau) \bigr]\bigr].
    \label{Bogotransf4}
\end{align}
In the following, we evaluate Eq.~\eqref{Bogotransf2} step by step.

\paragraph{Step 2: Trace over $\chi$.}

First, we evaluate the commutator:
\begin{align}
	\left[ \hat{T}_{S_2}(\tau_2), \hat{T}_{S_1}(\tau_1) \right] &\supset \left[ \hat{\chi}_{\mathbf{l_2}}(\tau_2) \hat{\chi}_{\mathbf{p_2}}(\tau_2), \hat{\chi}_{\mathbf{l_1}}(\tau_1) \hat{\chi}_{\mathbf{p_1}}(\tau_1) \right].
\end{align}
Using the symmetry of the dummy variables in the momentum integrals, we have:
\begin{align}
	\left[ \hat{\chi}_{\mathbf{l_2}}(\tau_2) \hat{\chi}_{\mathbf{p_2}}(\tau_2), \hat{\chi}_{\mathbf{l_1}}(\tau_1) \hat{\chi}_{\mathbf{p_1}}(\tau_1) \right] = 2 \left( \hat{\chi}_{\mathbf{l_2}}(\tau_2) \hat{\chi}_{\mathbf{l_1}}(\tau_1) + \hat{\chi}_{\mathbf{l_1}}(\tau_1) \hat{\chi}_{\mathbf{l_2}}(\tau_2) \right) \left[ \hat{\chi}_{\mathbf{p_2}}(\tau_2), \hat{\chi}_{\mathbf{p_1}}(\tau_1) \right].
\end{align}
The commutator part is written by the retarded Green function:
\begin{align}
	i a^2(\tau_2) \Theta(\tau_1 - \tau_2) [\hat{\chi}_{\mathbf{p_1}}(\tau_1), \hat{\chi}_{\mathbf{p_2}}(\tau_2)] = G^R_{p_1}(\tau_1, \tau_2) (2\pi)^3 \delta(\mathbf{p_1} + \mathbf{p_2}). \label{defretG}
\end{align}
The Keldysh Green function writes the operator part:
\begin{align}
	a^2(\tau_2) \, \text{Tr}\left[\hat{\varrho} (\hat{\chi}_{\mathbf{l_2}}(\tau_2) \hat{\chi}_{\mathbf{l_1}}(\tau_1) + \hat{\chi}_{\mathbf{l_1}}(\tau_1) \hat{\chi}_{\mathbf{l_2}}(\tau_2))\right]_\chi = G^K_{l_1}(\tau_1, \tau_2) (2\pi)^3 \delta(\mathbf{l_1} + \mathbf{l_2}). \label{defKelG}
\end{align}
With these Green functions, Eq.~\eqref{defemts} yields
\begin{align}
	\text{Tr}\left[\hat{\varrho} \left[ \hat{T}_{S_2}(\tau_2), \hat{T}_{S_1}(\tau_1) \right]\right]_\chi &= a^2(\tau_1) a^2(\tau_2) \int \frac{d^3l_1 d^3l_2 d^3p_1 d^3p_2}{(2\pi)^{12}} 
	\notag 
	\\
	&\times
	(2\pi)^3 \delta(\mathbf{k_1} + \mathbf{l_1} + \mathbf{p_1})(2\pi)^3 \delta(\mathbf{k_2} + \mathbf{l_2} + \mathbf{p_2}) \notag \\
	& \times e^{i_1 j_1(s_1)}_{\mathbf{k_1}} e^{i_2 j_2(s_2)}_{\mathbf{k_2}} l_{1i_1} p_{1j_1} l_{2i_2} p_{2j_2} \, 
	\notag 
	\\
	&\times
	\text{Tr}\left[\hat{\varrho} \left[ \hat{\chi}_{\mathbf{l_2}}(\tau_2) \hat{\chi}_{\mathbf{p_2}}(\tau_2), \hat{\chi}_{\mathbf{l_1}}(\tau_1) \hat{\chi}_{\mathbf{p_1}}(\tau_1)\right]\right]_\chi,
\end{align}
which reduces to:
\begin{align}
	&2i (2\pi)^3 \delta(\mathbf{k_1} + \mathbf{k_2}) \frac{a^2(\tau_1)}{a^2(\tau_2)} \int \frac{d^3l_1 d^3p_1}{(2\pi)^6}  (2\pi)^3 \delta(\mathbf{k_2} - \mathbf{l_1} - \mathbf{p_1}) e^{i_1 j_1(s_1)}_{\mathbf{k_1}} e^{i_2 j_2(s_2)}_{-\mathbf{k_1}} 
	\notag 
	\\
	&\times
	p_{1i_1} p_{1j_1} p_{1i_2} p_{1j_2} G^R_{p_1}(\tau_1, \tau_2) G^K_{l_1}(\tau_1, \tau_2),
\end{align}
where we integrate out $\mathbf{l_2}$ and $\mathbf{p_2}$, and used the transverse condition for the polarization tensors.

Now, consider $k \ll l \sim p$, i.e., the gravitational wave wavelength is sufficiently longer than that of thermal fields.
Then, we integrate out $\mathbf{l_1}$ and find:
\begin{align}
	&2i (2\pi)^3 \delta(\mathbf{k_1} + \mathbf{k_2}) \frac{a^2(\tau_1)}{a^2(\tau_2)} \int \frac{d\hat{p}}{4\pi} e^{i_1 j_1(s_1)}_{\mathbf{k_1}} e^{i_2 j_2(s_2)*}_{\mathbf{k_1}} \notag 
	\\
	&\times
	\hat{p}_{i_1} \hat{p}_{j_1} \hat{p}_{i_2} \hat{p}_{j_2} \int \frac{p^2 dp}{2\pi^2} p^4 G^K_{p}(\tau_1, \tau_2) G^R_{p}(\tau_1, \tau_2),
\end{align}
where $\hat{p} \equiv \mathbf{p}/p$.
The angular integral is evaluated as:
\begin{align}
	\int \frac{d\hat{p}}{4\pi} \hat{p}_{i_1} \hat{p}_{j_1} \hat{p}_{i_2} \hat{p}_{j_2} = \frac{1}{15} \left( \delta_{i_1 j_1} \delta_{i_2 j_2} + \delta_{i_1 j_2} \delta_{i_2 j_1} + \delta_{i_1 i_2} \delta_{j_2 j_1} \right),
\end{align}
which leads to
\begin{align}
	\int \frac{d\hat{p}}{4\pi} e^{i_1 j_1(s_1)}_{\mathbf{k_1}} e^{i_2 j_2(s_2)*}_{\mathbf{k_1}} \hat{p}_{i_1} \hat{p}_{j_1} \hat{p}_{i_2} \hat{p}_{j_2} = \frac{2}{15} \delta^{(s_1)(s_2)}.
\end{align}
To summarize, we obtain
\begin{align}
	\text{Tr}\left[\hat{\varrho} \left[ \hat{T}_{S_2}(\tau_2), \hat{T}_{S_1}(\tau_1) \right]\right]_\chi &= 2i \delta^{(s_1)(s_2)} (2\pi)^3 \delta(\mathbf{k_1} + \mathbf{k_2}) X(\tau_1,\tau_2) .\label{expTderive2}
\end{align}
The window function~$X$ is defined as
\begin{align}
	X(\tau_1,\tau_2) \equiv   \frac{2}{15} \frac{a^2(\tau_1)}{a^2(\tau_2)} \int \frac{p^2 dp}{2\pi^2} p^4 G^K_{p}(\tau_1, \tau_2) G^R_{p}(\tau_1, \tau_2).\label{expTderive:defX}
\end{align}
We provide a summary of necessary Green functions in appendix~\ref{appGreen}.
Using these results, Eq.~\eqref{expTderive:defX} is evaluated analytically:
\begin{align}
&X(\tau_1,\tau_2) = \frac{1}{15 \pi^2} \left[- \frac{3}{8 \Delta \tau^5} + \text{csch}^5\left(\frac{2 \pi \Delta \tau}{\beta}\right)  \right.
    \notag \\
    &  \left. \times \frac{\pi^5}{\beta^5} \left(11 \cosh\left(\frac{2 \pi \Delta \tau}{\beta}\right)
    + \cosh\left(\frac{6 \pi \Delta \tau}{\beta}\right)\right)  \right],\label{eq19}
\end{align}
which serves as a window function that peaks for $\Delta \tau = \tau_1 - \tau_2 \sim \beta/(2\pi) $.
We neglected the zero-temperature component when deriving Eq.~\eqref{eq19}.
As usual, the vacuum contribution may contain divergences, but these are distinct from finite-temperature effects, as the renormalization of physical constants at zero temperature is independent of the physics at finite temperature~\cite{Landsman:1986uw}.
The divergence for the finite-temperature part is avoided here because the thermal distribution suppresses contributions from high and low momentum modes.
Since $X$ is a window for a short interval $\tau_2 \lesssim \tau_1 \lesssim \tau_2 + \beta/2\pi$, we can assume that the graviton mode functions are approximately constant over this interval. This is nothing but the Markovian approximation, which allows us to carry out the $\tau_1$ integration straightforwardly.
In the high-temperature limit, $\beta / \tau \ll 1$, we find
\begin{align}
    \lim_{\beta / \tau \to 0} \int_{\tau_2}^\tau d\tau_1 \, X(\tau_1,\tau_2) = \frac{\pi^2}{450 \beta^4}.
\end{align}

\paragraph{Step 3: the trace with respect to gravitons.}

This step closely parallels the photon case, except that we must properly account for the time dependence of the number operator in an expanding universe.  
To do so, we express the graviton operator at $\tau_{1,2}$ in terms of Eq.~\eqref{def:instan}:
\begin{align}
    \hat{h}^{(s_i)}_{\mathbf k_i}(\tau_i) = u_{k_i}(\tau_i, \tau) \hat{d}^{(s_i)}_{\mathbf{k}_i} + u^*_{k_i}(\tau_i, \tau) \hat{d}^{(s_i)\dagger}_{-\mathbf{k}_i}.\label{modefunctioninst}
\end{align}
Here, $u$ denotes the positive-frequency mode function evaluated at $\tau_i$ relative to the instantaneous annihilation operator at $\tau$.  
Equation~\eqref{modefunctioninst} can be expanded in terms of the creation and annihilation operators defined at any time using the general Bogoliubov relation in Eq.~\eqref{instdefd}.  
We use Eq.~\eqref{modefunctioninst} because it simplifies the subsequent algebra.  
We then need only to expand the commutator:
\begin{align}
	&\hat{h}^{S_2}(\tau_2) \left[ \hat{h}^{S_1}(\tau_1), \hat{d}^{(s)\dagger}_{\mathbf{k}} \hat{d}^{(s)}_{\mathbf{k}} \right] 
	\notag \\
	&= 
	\left(u_{k_2}(\tau_2;\tau)\hat d^{(s_2)}_{\mathbf k_2}+ u^*_{k_2}(\tau_2;\tau)\hat d^{(s_2)\dagger}_{-\mathbf k_2}\right) 
	\left[\left(u_{k_1}(\tau_1;\tau)\hat d^{(s_1)}_{\mathbf k_1}+ u^*_{k_1}(\tau_1;\tau)\hat d^{(s_1)\dagger}_{-\mathbf k_1}\right), \hat{d}^{(s)\dagger}_{\mathbf{k}} \hat{d}^{(s)}_{\mathbf{k}} \right]
	\notag \\
	&= 
	\left(u_{k_2}(\tau_2;\tau)\hat d^{(s_2)}_{\mathbf k_2}+ u^*_{k_2}(\tau_2;\tau)\hat d^{(s_2)\dagger}_{-\mathbf k_2}\right) 
	\Bigl(u_{k_1}(\tau_1;\tau)\bigl[ \hat d^{(s_1)}_{\mathbf k_1},\hat{d}^{(s)\dagger}_{\mathbf{k}}\bigr]\hat{d}^{(s)}_{\mathbf{k}}  
	+ u^*_{k_1}(\tau_1;\tau)\hat{d}^{(s)\dagger}_{\mathbf{k}} \bigl[ \hat d^{(s_1)\dagger}_{-\mathbf k_1},\hat{d}^{(s)}_{\mathbf{k}} \bigr]\Bigr)
	\notag \\
	&= 
	u_{k_2}(\tau_2;\tau)\hat d^{(s_2)}_{\mathbf k_2} 
	\Bigl(u_{k_1}(\tau_1;\tau)\bigl[ \hat d^{(s_1)}_{\mathbf k_1},\hat{d}^{(s)\dagger}_{\mathbf{k}}\bigr]\hat{d}^{(s)}_{\mathbf{k}}  
	- u^*_{k_1}(\tau_1;\tau)\hat{d}^{(s)\dagger}_{\mathbf{k}} \bigl[\hat{d}^{(s)}_{\mathbf{k}} , \hat d^{(s_1)\dagger}_{-\mathbf k_1}\bigr]\Bigr)
	\notag \\
	&\quad+
	u^*_{k_2}(\tau_2;\tau)\hat d^{(s_2)\dagger}_{-\mathbf k_2} 
	\Bigl(u_{k_1}(\tau_1;\tau)\bigl[ \hat d^{(s_1)}_{\mathbf k_1},\hat{d}^{(s)\dagger}_{\mathbf{k}}\bigr]\hat{d}^{(s)}_{\mathbf{k}}  
	- u^*_{k_1}(\tau_1;\tau)\hat{d}^{(s)\dagger}_{\mathbf{k}} \bigl[ \hat{d}^{(s)}_{\mathbf{k}}, \hat d^{(s_1)\dagger}_{-\mathbf k_1} \bigr]\Bigr)
	\notag \\
	&= 
	 u_{k_1}(\tau_1;\tau)u_{k_2}(\tau_2;\tau)\hat d^{(s_2)}_{\mathbf k_2} \hat{d}^{(s)}_{\mathbf{k}} \bigl[ \hat d^{(s_1)}_{\mathbf k_1},\hat{d}^{(s)\dagger}_{\mathbf{k}}\bigr]
	 - u^*_{k_1}(\tau_1;\tau)u_{k_2}(\tau_2;\tau)\hat d^{(s_2)}_{\mathbf k_2} \hat{d}^{(s)\dagger}_{\mathbf{k}} \bigl[\hat{d}^{(s)}_{\mathbf{k}} , \hat d^{(s_1)\dagger}_{-\mathbf k_1}\bigr]
	\notag \\
	&\quad+
	u_{k_1}(\tau_1;\tau)u^*_{k_2}(\tau_2;\tau)\hat d^{(s_2)\dagger}_{-\mathbf k_2} \hat{d}^{(s)}_{\mathbf{k}} \bigl[ \hat d^{(s_1)}_{\mathbf k_1},\hat{d}^{(s)\dagger}_{\mathbf{k}}\bigr]
	 - u^*_{k_1}(\tau_1;\tau)u^*_{k_2}(\tau_2;\tau)\hat d^{(s_2)\dagger}_{-\mathbf k_2} \hat{d}^{(s)\dagger}_{\mathbf{k}} \bigl[ \hat{d}^{(s)}_{\mathbf{k}}, \hat d^{(s_1)\dagger}_{-\mathbf k_1}\bigr]
	\notag \\
	&= 
	 u_{k}(\tau_1;\tau)u_{k}(\tau_2;\tau)\hat d^{(s_2)}_{\mathbf k_2} \hat{d}^{(s)}_{\mathbf{k}} \bigl[ \hat d^{(s_1)}_{\mathbf k_1},\hat{d}^{(s)\dagger}_{\mathbf{k}}\bigr]   
	 - u^*_{k}(\tau_1;\tau)u^*_{k}(\tau_2;\tau)\hat d^{(s_2)\dagger}_{-\mathbf k_2} \hat{d}^{(s)\dagger}_{\mathbf{k}} \bigl[ \hat{d}^{(s)}_{\mathbf{k}}, \hat d^{(s_1)\dagger}_{-\mathbf k_1}\bigr]
	\notag \\
	&\quad+
	u_{k}(\tau_1;\tau)u^*_{k}(\tau_2;\tau)\hat d^{(s_2)\dagger}_{-\mathbf k_2} \hat{d}^{(s)}_{\mathbf{k}} \bigl[ \hat d^{(s_1)}_{\mathbf k_1},\hat{d}^{(s)\dagger}_{\mathbf{k}}\bigr]     
	 - u^*_{k}(\tau_1;\tau)u_{k}(\tau_2;\tau)\hat d^{(s_2)}_{\mathbf k_2} \hat{d}^{(s)\dagger}_{\mathbf{k}} \bigl[\hat{d}^{(s)}_{\mathbf{k}} , \hat d^{(s_1)\dagger}_{-\mathbf k_1}\bigr].
\end{align}

The following expectation values in the squeezed vacuum state are:
\begin{align}
	\langle \psi |\hat d^{(s_2)}_{\mathbf k_2} \hat{d}^{(s)}_{\mathbf{k}} |\psi \rangle 
	&=\langle \psi |\bigl(\mu_{k_2}\hat d^{(s_2)}_{\psi,\mathbf k_2} + \nu_{k_2}\hat d^{(s_2)\dagger}_{\psi, -\mathbf k_2}\bigr)  
	\bigl( \mu_k \hat{d}^{(s)}_{\psi, \mathbf{k}} + \nu_k \hat{d}^{(s)\dagger}_{\psi, - \mathbf{k}} \bigr)|\psi \rangle 
	= \mu_{k} \nu_k \bigl[ \hat d^{(s_2)}_{\psi,\mathbf k_2}  ,  \hat{d}^{(s)\dagger}_{\psi, - \mathbf{k}} \bigr],
	\\
	\langle \psi |\hat d^{(s_2)\dagger}_{-\mathbf k_2} \hat{d}^{(s)\dagger}_{\mathbf{k}} |\psi \rangle 
	&=\langle \psi |\bigl(\mu^*_{k_2}\hat d^{(s_2)\dagger}_{\psi,-\mathbf k_2} + \nu^*_{k_2}\hat d^{(s_2)}_{\psi, \mathbf k_2}\bigr)  
	\bigl( \mu_k^* \hat{d}^{(s)\dagger}_{\psi, \mathbf{k}} + \nu_k^* \hat{d}^{(s)}_{\psi, - \mathbf{k}} \bigr)|\psi \rangle 
	= \mu^*_{k} \nu^*_k \bigl[ \hat d^{(s_2)}_{\psi,\mathbf k_2}  ,  \hat{d}^{(s)\dagger}_{\psi,  \mathbf{k}} \bigr], 
	\\
	\langle \psi| \hat d^{(s_2)\dagger}_{-\mathbf k_2} \hat{d}^{(s)}_{\mathbf{k}}  | \psi \rangle 
	&= \langle \psi| \bigl( \mu^*_{k_2} \hat d^{(s_2)\dagger}_{\psi, -\mathbf k_2} +\nu^*_{k_2} \hat d^{(s_2)}_{\psi, \mathbf k_2}  \bigr)\bigl( \mu_k \hat{d}^{(s)}_{\psi, \mathbf{k}} + \nu_k \hat{d}^{(s)\dagger}_{\psi, -\mathbf{k}}\bigr)  | \psi \rangle
	= |\nu_k|^2 \bigl[\hat d^{(s_2)}_{\psi, \mathbf k_2}, \hat{d}^{(s)\dagger}_{\psi, -\mathbf{k}}\bigr],
	\\
	\langle \psi| \hat d^{(s_2)}_{\mathbf k_2} \hat{d}^{(s)\dagger}_{\mathbf{k}}| \psi \rangle 
	&= \langle \psi| \bigl( \mu_{k_2} \hat d^{(s_2)}_{\psi, \mathbf k_2} +\nu_{k_2} \hat d^{(s_2)\dagger}_{\psi, - \mathbf k_2}  \bigr)\bigl( \mu^*_k \hat{d}^{(s)\dagger}_{\psi, \mathbf{k}} + \nu^*_k \hat{d}^{(s)}_{\psi, -\mathbf{k}}\bigr)  | \psi \rangle
	= |\mu_k|^2 \bigl[ \hat d^{(s_2)}_{\psi, \mathbf k_2}, \hat{d}^{(s)\dagger}_{\psi, \mathbf{k}}\bigr].
\end{align}

Since $\mathbf k_1$ and $\mathbf k_2$ are symmetric dummy variables, symmetry implies that in the state $|\psi \rangle$ we obtain:
\begin{align}
	\langle \psi|\hat{h}^{S_2}(\tau_2)\bigl[ \hat{h}^{S_1}(\tau_1), \hat{d}^{(s)\dagger}_{\mathbf{k}} \hat{d}^{(s)}_{\mathbf{k}}\bigr]|\psi\rangle
	&= \delta^{ss_2}\,\delta^{ss_1}\,(2\pi)^3\,\delta(\mathbf k_1 -\mathbf k)\,(2\pi)^3\,\delta(\mathbf k_2 +\mathbf k)
	\notag \\
	&\quad\times 
	\bigl[
	u_{k}(\tau_1;\tau)\,u_{k}(\tau_2;\tau)\,\mu_k \nu_k
	- u^*_{k}(\tau_1;\tau)\,u^*_{k}(\tau_2;\tau)\,\nu^*_k \mu^*_k 
	\notag \\
	&\quad\quad
	+ u_{k}(\tau_1;\tau)\,u^*_{k}(\tau_2;\tau)\,|\nu_k|^2
	- u^*_{k}(\tau_1;\tau)\,u_{k}(\tau_2;\tau)\,|\mu_k|^2
	\bigr]. 
	\label{comh}
\end{align}
One may compute the analogous term $\bigl[\hat{h}^{S_1}(\tau_1), \hat{d}^{(s)\dagger}_{\mathbf{k}} \hat{d}^{(s)}_{\mathbf{k}}\bigr]\hat{h}^{S_2}(\tau_2)$ similarly; this operation effectively exchanges the factors $|\mu_k|^2$ and $|\nu_k|^2$ in the final two terms.

\paragraph{Summary.}

After a careful handling of the polarization sums and momentum integrations, one recovers the graviton analogue of Eq.~\eqref{dN2stim}:
\begin{align}
	{\rm Tr}&\bigl[\hat \varrho \,\tfrac{\hat{N}_{{\rm stim},\mathbf k}^{(s)}}{V}\bigr]
	= \Re\bigl[\zeta_k (2 |\nu_{k}|^2+1) + 2\sigma_k \mu_{k}^* \nu_{k}^*\bigr],
	\label{stimgrv}\\
    \zeta_k &\equiv 2i  \int^\tau d\tau_2 \int^\tau_{\tau_2} d\tau_1 \, u_k(\tau_2, \tau)\,u^*_k(\tau_1, \tau)\,\frac{X(\tau_1, \tau_2)}{M_{\rm pl}^2}, 
    \label{lambdaq}\\
    \sigma_k &\equiv 2i  \int^\tau d\tau_2 \int^\tau_{\tau_2} d\tau_1 \, u_k^*(\tau_2, \tau)\,u^*_k(\tau_1, \tau)\,\frac{X(\tau_1, \tau_2)}{M_{\rm pl}^2},
    \label{nuq}
\end{align}
Here, we have reordered the time integrals for convenience.  
The relative importance of stimulated emission in each frequency bin is then characterized by
\begin{align}
	r_k \equiv  2\Re\Bigl[\zeta_k+ \tfrac{2\sigma_k \mu_{k}^* \nu_{k}^*}{2|\nu_k|^2+1}\Bigr].
	\label{def:r}
\end{align}

The stimulated emission rate $r_k$ enters the observable energy density via the graviton number operator:
\begin{align}
\rho_{\rm GW} = \frac{1}{a^4 V}\int \frac{d^3k}{(2\pi)^3} k \sum_s \langle \psi| \hat N^s_{\mathbf k}|\psi\rangle.
	\label{defrhogw}
\end{align}
One then defines the dimensionless GW energy spectrum by $\rho_{\rm GW}/\rho_{\rm tot} = \int d\ln k\,\Omega_{\rm GW}(k)$, with $\rho_{\rm tot}$ the energy density of $\chi$ in our setup.  In this decomposition, the stimulated contribution is simply
\begin{align}
	  \Omega^{\rm stim}_{\rm GW}(k) = r_k\,\Omega^{\rm free}_{\rm GW}(k),
\end{align}
implying that $r_k$ can in principle be extracted from observations of the stochastic GW background.

\section{Minkowski background}
\label{sec:minkowski}

We now evaluate the cosmological stimulated emission rate $r_k$ in a thermal radiation bath.  
For simplicity, we start with a Minkowski background, setting $a=1$ and treating $\chi$ as a spectator thermal field.  
In this limit, comoving variables coincide with physical ones.  
The positive-frequency mode function in Eq.~\eqref{modefunctioninst} is
\begin{align}
	u_k(\tau_1, \tau) = \frac{e^{-i k (\tau_1 - \tau)}}{\sqrt{2k}}.
\end{align}
Since no particle production occurs dynamically in Minkowski space, the vacuum $|0\rangle$ is uniquely defined (up to a phase).  
As our initial state, we prepare a squeezed state as an excited state: $ |\psi\rangle \equiv \hat{S}\,|0 \rangle,$
using a non dynamical Bogoliubov transformation to mimic a cosmological scenario.  We set  
\begin{align}
	|\nu|^2 = n_0, \quad
	\mu\nu = e^{i\theta_0} \sqrt{n_0(1+n_0)}, \label{inimin}
\end{align}
where $n_0$ is the initial graviton occupancy and $\theta_0$ is the squeezing phase.  
Note that $\theta_0$ controls the squeezing amplitude; we have omitted any overall unphysical phase.  
Without interaction, $n_0$ remains constant and $\mu\nu\propto e^{-2ik\tau}$.  
Evaluating Eqs.~\eqref{lambdaq} and \eqref{nuq} in this setup yields
\begin{align}
	\zeta_k = \frac{i \pi ^2 \tau e^{\frac{i k\beta}{2 \pi }} }{450 k M_{\rm pl}^2 \beta^4}, \quad
	\sigma_k = \frac{i \pi ^2 e^{-i k\tau}  \sin(k\tau)}{450 k^2 M_{\rm pl}^2 \beta^4},
\end{align}
in the high-temperature limit.  Hence the emission rate becomes
\begin{align}
    r_k = -\frac{\pi ^2 \sqrt{1 + \frac{1}{n_0}} \sin(k \tau) \sin(k\tau - \theta_0)}{225 M_{\rm pl}^2 k^2 \beta^4} + \mathcal{O}(\beta^{-3}). \label{rmin}
\end{align}
The sign of $r_k$ depends on $\theta_0$, allowing either net emission or absorption; the effect grows for longer wavelengths and higher temperatures.  
Although the thermal bath itself has no population inversion, the graviton squeezed vacuum behaves as an effectively inverted sector relative to the instantaneous vacuum.  
Since ${\rm Re}[\zeta_k]=\mathcal O(\beta^{-3})$, $r_k$ is Planck mass suppressed for graviton number eigenstates $|N\rangle\langle N|$.  
Thus, the squeezing correlation $\theta_0$ couples to graviton production, analogous to particle creation in curved spacetime.

\medskip

\begin{figure}
	\centering
	\includegraphics[width=0.7\linewidth]{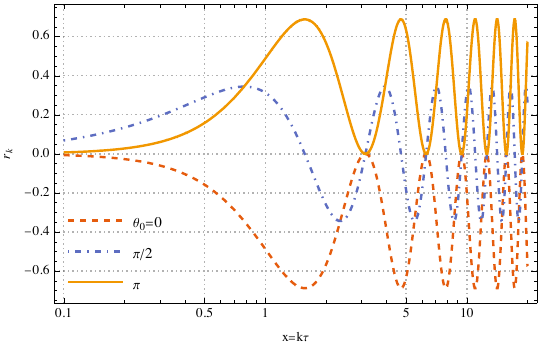}
	\caption{The stimulated emission rate~\eqref{rmin} for $n_0=1$ as functions of $x=k\tau$. Thermal bath temperature and graviton wavelength are set to $T=0.1$ GeV and $10^6$ m. The amplitude varies as Eq.~\eqref{ampdef}. The initial phases are set to $\theta_0=0,\pi/2,$ and $\pi$, and these are expressed by orange dashed, blue dot-dashed and yellow solid lines. The net emission rate can be nonzero.}
	\label{fig1}
\end{figure}

Cosmological stimulated emission may be relevant for ground-based GW detectors such as LIGO/Virgo~\cite{LIGOScientific:2016aoc}.  
Observed GW frequencies are typically $\mathcal{O}(100)\,\text{Hz}$, corresponding to wavelengths $\lambda=2\pi/k\sim\mathcal{O}(10^6)\,\text{m}$.  
Although laboratory temperatures are limited, these wavelengths are enormous in Planck units.  Equation~\eqref{rmin} can be recast as
\begin{align}
    r_k \sim \left( \frac{\lambda}{10^6 \, \text{m}} \right)^2 \left( \frac{T}{0.1 \, \text{GeV}} \right)^4,\label{ampdef}
\end{align}
where $T=\beta^{-1}$.
Hence, for $\beta^{-1}\gg \mathcal{O}(0.1)\,\text{GeV}$, $r_k$ may become significant.  
Such energy scales are extremely high but still below the Planck scale and within reach of modern accelerators (e.g.\ LHC at 7\,TeV).  
Although a detailed quantitative study is beyond this work, investigating cosmological stimulated emission further could prove important.
We present the emission rate for the various initial phases $\theta_0$ in Fig.~\ref{fig1}, which indicates both emission and absorption, depending on the initial condition.

However, one cannot take arbitrarily high temperatures here due to perturbativity in $\hat{H}_I$ and backreaction of the thermal bath on the background.  
Identifying $1/(2\tau)$ with the Hubble parameter, radiation backreaction on flat space becomes significant if $\beta^{-1} \gtrsim \sqrt{M_{\rm pl}/(2\tau)}\sim\sqrt{kM_{\rm pl}}$.  
Therefore, our Minkowski analysis is reliable provided $r_k \lesssim 0.01$.

\section{Radiation-dominated universe}\label{sec:rad‐dom}
The above analysis of the Minkowski background suggests that stimulated emission in a radiation-dominated universe is essential for higher temperatures and longer wavelengths.
In cosmological scenarios, the initial quantum state is given by inflation instead of Eq.~\eqref{inimin}.
Without interaction, one can integrate the field equations from the remote past during inflation to $\tau$ during radiation dominant.
With a careful connection of mode functions at reheating we can write the Bogoliubov coefficients in Eq.~\eqref{stimgrv}:
\begin{align}
  |\nu_k|^2 &= \frac{1}{8 k^6 \tau^2 \tau_R^4}\Bigg[1 + 2 k^4 \tau_R^4 + 2 k^2 \tau^2 
  \notag 
  \\
  &
  + 2 k \left( 2 k^2 \tau \tau_R^2 - \tau + \tau_R \right) \sin\left( 2 k (\tau - \tau_R) \right) 
  \notag 
  \\
  &
  + \left( 2 k^2 \tau_R \left( \tau_R - 2 \tau \right) - 1 \right) \cos\left( 2 k (\tau - \tau_R) \right) \Bigg], 
  \label{nunu} \\
  \mu_k \nu_k &= \frac{1}{8 k^6 \tau^2 \tau_R^4}\Bigg[ -i (2 k \tau - i) \left( 2 k^4 \tau_R^4 + 1 \right) 
  \notag 
  \\
  &
  + \left( 4 k^4 \tau^2 \tau_R^2 + 4 i k^3 \tau \tau_R (\tau - \tau_R) 
  \right.
  \notag 
  \\
  &
  \left.- 2 k^2 (\tau - \tau_R)^2 + 2 i k \tau + 1 \right) \cos\left( 2 k (\tau - \tau_R) \right) 
  \notag 
  \\
  &+ 2 k \left( - \tau_R + \tau \left( k \tau - i \right) 
  \right.
  \notag 
  \\
  &\left.\times \left( 2 k \tau_R \left( 1 - i k \tau_R \right) + i \right) \right) \sin\left( 2 k (\tau - \tau_R) \right)\Bigg],
  \label{numu}
\end{align}
where $\tau_R$ is the reheating time, and instantaneous reheating is considered.
These coefficients are found in Eq.~\eqref{defbt2}; see Appendix~\ref{sumofmodefunc} for details.
The positive frequency mode function in Eq.~\eqref{modefunctioninst}, with scale factor $a(\tau) = \tau /( H \tau_{\rm R}^2)$, is given by  
\begin{align}
    u_k(\tau_1,\tau) = \frac{ H  \tau_{R}^2 \left(\sin (k (\tau_1 -\tau))+k \tau  e^{-i k (\tau_1 -\tau)}\right)}{\sqrt{2} k^{3/2} \tau  \tau_{1}}. \label{mdinst}
\end{align}
This equation is found in Eq.~\eqref{defmdinsthen}.
Here, $H$ is the inflationary Hubble parameter, as instantaneous reheating is assumed.  
With this setup, we evaluate Eqs.~\eqref{lambdaq} and \eqref{nuq}.  

\medskip
We derived an analytic expression for $r_k$, but it is tedious.
The full expression is found in appendix~\ref{appC}; here, we discuss its asymptotic expression.
We are interested in the limit $\tau/\tau_R \to \infty$, i.e., the spectrum of gravitons sufficiently after reheating as we measure it in the end.
In this limit, $r_k$ simplifies to 
\begin{align}
\lim_{\tau/\tau_R\to \infty } r_k =f(k\tau_R),  \label{eqrkflrw}
\end{align}
where we defined
\begin{align}
	f(x_R)&\equiv \frac{2}{5} \big[2 \text{Ci}(2 x_R) \left(\left(2 x_R^2-1\right) \cos (2 x_R)-2 x_R \sin (2 x_R)\right)
	\notag 
	\\
	&-(\pi -2 \text{Si}(2 x_R)) \left(\left(2 x_R^2-1\right) \sin (2 x_R)+2 x_R \cos (2 x_R)\right)+2\big].
\end{align}

We can see different features in $k\tau_R\ll1$ and $k\tau_R\gg1$.
Graviton modes well inside the horizon at reheating time, i.e., the modes remain inside the horizon throughout yield 
\begin{align}
	\lim_{k\tau_R\to \infty }\left( \lim_{\tau/\tau_R\to \infty} r_k \right)= -\frac{2}{5}.
\end{align}
Thus, $r_k$ stays in the perturbative regime, $|r_k|<1$, and the stimulated absorption is observed.
Note that $M_{\rm pl}^2$ in the denominator is canceled by using the Friedmann equation: $3 M_{\rm pl}^2 H^2 = \rho_\chi (\tau_R)$, we find $\tau^4_{\rm R}/\beta^4 = 90 M_{\rm pl}^2 / (\pi^2 H^2)$, assuming the radiation is dominated by $\chi$.
These modes never exit the horizon, so the subtle issues associated with super-horizon modes do not arise.

\begin{figure}
\centering
	\includegraphics[width = .7 \linewidth]{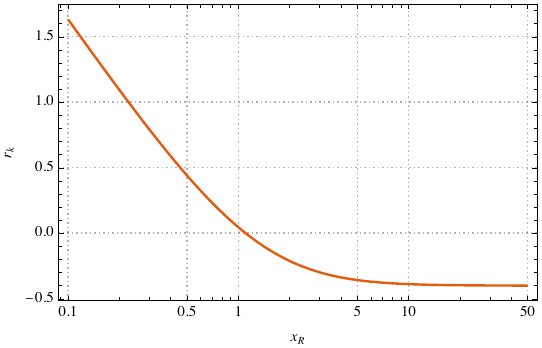}
	\caption{The stimulated emission rate~\eqref{eqrkflrw} in the $\tau/\tau_R\to \infty$ limit as a function of the momentum scale at reheating time. This limit is taken by fixing $k$. $x_R = k\tau_R > 1$ indicates modes initially inside the horizon, while $x_R \leq 1$ corresponds to superhorizon modes at reheating. The subhorizon mode is reduced by a factor of 40\% in the one-loop analysis in the present setup and is expected to remain in the perturbative regime. The perturbative analysis completely fails for modes with $x_R < 0.2$.}

	\label{fig2}
\end{figure}

On the other hand, the graviton mode initially super horizon at $\tau_R$ yields
\begin{align}
	\lim_{k\tau_R\to 0}\left( \lim_{\tau/\tau_R\to \infty} r_k \right) = \frac{4}{5} \Big( 1 - \gamma_{\rm E}  - \ln(2 k\tau_R) \Big),\label{stimsuperhorizon}
\end{align}
where $\gamma_{\rm E} = 0.577216$ is the Euler-Mascheroni constant.
Eq.~\eqref{eqrkflrw} is also displayed in Fig.~\ref{fig2}.

Eq.~\eqref{stimsuperhorizon} exceeds unity for $k\tau_R \lessapprox 1$.
Sensitivity to longer wavelength modes is a common feature in stimulated emission for quantum electrodynamics.
However, the longest wavelength is bounded by the atomic energy gap in that case.
Stimulated emission caused by the squeezed vacuum state in the radiation-dominant universe is not constrained in this way.
The large stimulated emission observed at 1-loop order suggests a breakdown in the perturbative approach, an issue commonly encountered in thermal field theory~\cite{Laine:2016hma}, which would need to be incorporated into a more comprehensive analysis to be addressed in future work.
While Eq.~\eqref{stimsuperhorizon} does not depend on $\tau$, the log dependence on the initial time $\tau_R$ implies a sort of secular growth, a common issue in the initial value problem in quantum field theory at late time, which might be resumed or removed in the end~\cite{Boyanovsky:2003ui,Iso:2017hvi}. The stimulated emission occurs in a short time interval $\Delta \tau \sim \beta$ as indicated by the window function~\eqref{eq19}. Therefore, the secular growth does not arises due to the ignorance of dissipation in the free scalar field model. 

While it depends on the reheating time, $k\tau_R \gg1$ is quite small scale, so $k\tau_R \ll 1$ concerns more. 
As described above, one cannot quantitatively rely on Eq.~\eqref{stimsuperhorizon}.
However, assuming there exists such an effect even after proper prescriptions, it has tremendous implications for cosmological observations. The amplitudes of primordial perturbations are directly related to the size of the slow-roll parameter and the inflationary Hubble scale~\cite{Planck:2018jri}. The present constraints on the primordial gravitational wave amplitude become effectively stronger, provided that the net stimulated emission is positive and vice versa.

\section{Conclusions}\label{sec:conclusion}

In this paper, we have considered the evolution of gravitons in a squeezed vacuum state immersed in a thermal radiation bath, motivated by cosmological gravitons in a radiation-dominated universe.  Naively, one might expect either simple absorption of gravitons in the thermal bath or no observable effect due to Planck-scale suppression of the interaction.  However, we found that, depending on the squeezed parameters, the net process can be either emission or absorption.  We interpret this phenomenon as stimulated emission of gravitons, analogous to the mechanism behind lasers in quantum electrodynamics.

The efficiency of graviton stimulated emission is characterized by $T^4/(k M_{\rm pl})^2$, with $T$ the thermal bath temperature, $k$ the infrared graviton momentum, and $M_{\rm pl}$ the reduced Planck mass.  Thus, longer-wavelength modes and higher temperatures lead to stronger stimulated effects.  In Minkowski spacetime, the thermal bath acts as a spectator, and $T$ cannot be arbitrarily high: as $T$ grows for a given $k$, backreaction on the background spacetime becomes non-negligible, motivating our focus on the radiation-dominated universe.  Unlike laboratory lasers, cosmological scenarios impose no lower bound on IR graviton momenta, since inflation generates a nearly scale-invariant spectrum.  Moreover, the Friedmann equation $T \sim \sqrt{M_{\rm pl} H}$ implies the efficiency scales as $H^2/k^2$.  Thus, our perturbative estimation fails on superhorizon scales ($H/k>1$), leading to secular growth.  We leave a detailed prescription for IR secular growth to future work, but note that the effect remains perturbative in the sub-horizon limit, as suggested by the Minkowski result.  We found the spectrum is reduced by 40\% in the one-loop analysis for the graviton modes inside the horizon at reheating.

As a thought experiment, we proposed placing a LIGO/Virgo-like detector in a thermal plasma.  By setting the plasma temperature high, one expects an enhancement of the graviton signal, potentially serving as a mechanism to amplify the signal in gravitational-wave observations.  In the Minkowski background, the typical emission rate was estimated as $r_k \sim (\lambda/10^6\text{m})^2 (T/0.1\,\text{GeV})^4$ with $\lambda\sim k^{-1}$.  Although 0.1\,GeV is extremely high, it is not inaccessible to humanity, as the LHC has already achieved 7\,TeV in instantaneous collisions.  Exploring this effect in high-temperature astrophysical phenomena, such as ultra-high-energy cosmic-ray sources, could be interesting.

This analysis includes several simplifying assumptions that capture the primary physical implications for an initial study.  Specifically, thermal radiation is modeled by a massless free scalar field with an initial thermal distribution, neglecting any self-interactions of $\chi$.  A more realistic treatment involving Standard Model fields would introduce characteristic diffusion scales and damping effects.  Nevertheless, since stimulated emission occurs over an extremely short interval, $\Delta \tau \sim \beta$, which is the fastest scale in the system, dissipative effects introduced by realistic thermal fields are expected to be the sub-leading effect. For cosmological radiation in strong coupling regime, such as the quark-gluon plasma, more detailed analysis including the mean-free-path scale will be required.  We also note that cosmological neutrinos after decoupling are well described by a collisionless thermal bath of massless particles.  Here, we ignore tensor perturbations in $\beta$, which modify the linear solution and result in damping of tensor modes~\cite{Weinberg:2003ur}, potentially offsetting the enhancement.

Finally, in a separate project the author and collaborators extend the one-loop in–in formalism to Heisenberg-evolution equations averaged over general quantum states~\cite{Ota:2025yeu}, and will compare Boltzmann-equation approaches with the one-loop analysis to investigate the convergence and validity of the perturbative expansion.

\appendix

\acknowledgments
The author would like to thank Misao Sasaki, Ryo Saito, Tsutomu Yanagida, Jun'ichi, Yokoyama, Yi Wang, and Richard Woodard for useful discussions.
The author would like to thank Yuhang Zhu for careful reading of our manuscript.
This work was supported in part by the National Natural Science Foundation of China under Grant No. 12347101 and 12403001.

\section{Various Mode Functions}\label{sumofmodefunc}

This section provides the explicit formulas for the mode functions in various backgrounds and vacuum choices.

\subsection{Fundamental Matrix} 
Let us begin by solving the dynamics of gravitons. The Hamiltonian equations are given by
\begin{align}
	\frac{\delta H_h}{\delta h^i{}_j} &= -\pi'^j{}_i =  -a^2 \partial^2 h^j{}_i, \\
	\frac{\delta H_h}{\delta \pi^i{}_j} &= h'^j{}_i = \frac{\pi^j{}_i}{a^2},
\end{align}
which, in Fourier space, can be expressed by the following matrix equation:
\begin{align}
	Y' = M Y, \label{fundaeom}
	~
	Y \equiv 
	\begin{pmatrix}
		h^{(s)}_{\mathbf k} \\
		\pi^{(s)}_{\mathbf k}
	\end{pmatrix},
	~
	M \equiv
	\begin{pmatrix}
		0 & \frac{1}{a^2} \\
		- a^2 k^2 & 0
	\end{pmatrix}.
\end{align}
We have suppressed the indices $\mathbf k$ and $s$ for notational simplicity.

Let us introduce the fundamental matrix $\mathcal{O}(\tau,\tau_0)$, defined by
\begin{align}
	\mathcal{O}(\tau,\tau_0)' &= M \mathcal{O}(\tau,\tau_0), \quad \mathcal{O}(\tau_0,\tau_0) =
	\begin{pmatrix}
		1 & 0 \\
		0 & 1
	\end{pmatrix}.
\end{align}
Using the fundamental matrix, we find the solution for a given initial condition $Y(\tau_0)$ as
\begin{align}
	Y(\tau) = \mathcal{O}(\tau,\tau_0) Y(\tau_0).
\end{align}
For a given background solution $a$, the fundamental matrix is found as follows. By comparing the components of Eq.~\eqref{fundaeom}, one finds
\begin{align}
	-a^2 k^2 O_{11} &= (a^2 O_{11}')', \\
	-a^2 k^2 O_{12} &= (a^2 O_{12}')'.
\end{align}
We solve these differential equations with the initial conditions
\begin{align}
	O_{11}(\tau_0) &= 1, \quad O_{11}'(\tau_0) = \frac{O_{21}(\tau_0)}{a(\tau_0)^2} = 0, \\
	O_{12}(\tau_0) &= 0, \quad O_{12}'(\tau_0) = \frac{O_{22}(\tau_0)}{a(\tau_0)^2} =  \frac{1}{a(\tau_0)^2}.
\end{align}
Then, we find the rest of the components by computing
\begin{align}
	O_{21} &= a^2 O_{11}', \\
	O_{22} &= a^2 O_{12}'.
\end{align}
Once we find the fundamental matrix, we can write the mode functions and the Bogoliubov transformations systematically.

Using the \textit{instantaneous} annihilation and creation operators
\begin{align}
	\binom{\hat{d}_{\mathbf{k}}}{\hat{d}^\dagger_{-\mathbf{k}}} \equiv Q_k \binom{\hat{h}^{(s)}_{\mathbf{k}}}{\hat{\pi}^{(s)\dagger}_{\mathbf{k}}}, ~
	Q_k \equiv \begin{pmatrix}
		a(\tau)\sqrt{\frac{k}{2}} & \frac{i}{a(\tau)\sqrt{2k}} \\
		a(\tau)\sqrt{\frac{k}{2}} & -\frac{i}{a(\tau)\sqrt{2k}}
	\end{pmatrix},
\end{align}
we write
\begin{align}
	Y(\tau) &= \mathcal{O}(\tau,\tau_0) Q_k(\tau_0)^{-1} \binom{\hat d_{\mathbf k}(\tau_0)}{\hat d^\dagger_{-\mathbf k}(\tau_0)}, \\
	\binom{\hat d_{\mathbf k}(\tau)}{\hat d^\dagger_{-\mathbf k}(\tau)} &= Q_k(\tau) \mathcal{O}(\tau,\tau_0) Q_k(\tau_0)^{-1} \binom{\hat d_{\mathbf k}(\tau_0)}{\hat d^\dagger_{-\mathbf k}(\tau_0)}.
\end{align}
Hence, we find
\begin{align}
	\begin{pmatrix}
	u_k(\tau,\tau_0) & u^*_k(\tau,\tau_0) \\
	v_k(\tau,\tau_0) & v^*_k(\tau,\tau_0)
	\end{pmatrix}
	&= \mathcal{O}(\tau,\tau_0) Q_k(\tau_0)^{-1}, \\
	\begin{pmatrix}
	\mu_k(\tau,\tau_0) & \nu_k(\tau,\tau_0) \\
	\nu_k^*(\tau,\tau_0) & \mu_k^*(\tau,\tau_0)
	\end{pmatrix}
	&= Q_k(\tau) \mathcal{O}(\tau,\tau_0) Q_k(\tau_0)^{-1}. \label{defbt}
\end{align}

\subsection{Connection of Solutions} 
When connecting two different solutions, $\mathcal{O}(\tau,\tau_\star)$ and $\mathcal{O}(\tau_\star,\tau_0)$, at $\tau_\star$, we simply multiply them:
\begin{align}
	\mathcal{O}(\tau,\tau_0) = \mathcal{O}(\tau,\tau_\star)\mathcal{O}(\tau_\star,\tau_0).
\end{align}
As an example, consider inflation followed by a radiation-dominant phase. The inflationary phase ends at $-\tau_{\rm R} < 0$, and then there is an instantaneous transition to radiation dominance at $\tau_{\rm R} > 0$. The scale factor and its derivative, i.e., the Hubble parameter, are continuous at the transition for
\begin{align}
	a(\tau) = 
	\begin{cases}
		-\frac{1}{ H \tau}, & \quad \tau < -\tau_{\rm R} < 0, \\
		\frac{1}{ H \tau_{\rm R}}\left( \frac{\tau}{\tau_{\rm R}}\right), & \quad 0 < \tau_{\rm R} < \tau.
	\end{cases} \label{scalefactora}
\end{align}
where $ H $ is the Hubble parameter at $\tau = \tau_{\rm R}$. Given the scale factors, the fundamental matrices for radiation dominance, $\mathcal{O}_{\rm RD}$, and inflation, $\mathcal{O}_{\rm dS}$, will be found. We then find the field operator during radiation dominance as
\begin{align}
	\hat{Y}(\tau) &= \mathcal{O}_{\rm RD}(\tau,\tau_{\rm R}) \hat{Y}(\tau_{\rm R}) 
	= \mathcal{O}_{\rm RD}(\tau,\tau_{\rm R}) \mathcal{O}_{\rm dS}(-\tau_{\rm R},\tau_0) \hat{Y}(\tau_0). \label{connect:sol}
\end{align}
From Eq.~\eqref{connect:sol}, the mode functions during radiation dominance with respect to the inflationary adiabatic vacuum are written as
\begin{align}
	\begin{pmatrix}
		u^{\rm dS \to RD}_k(\tau) & u^{\rm dS \to RD*}_k(\tau) \\
		v^{\rm dS \to RD}_k(\tau) & v^{\rm dS \to RD*}_k(\tau)
	\end{pmatrix}
	= \lim_{\tau_0 \to -\infty} \mathcal{O}_{\rm RD}(\tau,\tau_{\rm R}) \mathcal{O}_{\rm dS}(-\tau_{\rm R},\tau_0) Q_k^{-1}(\tau_0). \label{dstordmd}
\end{align}

\subsection{Fundamental Matrices in Various Backgrounds}

In a Minkowski background, where $a = 1$, we find
\begin{align}
	O_{11}(\tau,\tau_0) &= \cos k(\tau - \tau_0), \\
	O_{12}(\tau,\tau_0) &= \frac{1}{k}\sin k(\tau - \tau_0), \\
	O_{21}(\tau,\tau_0) &= -k \sin k(\tau - \tau_0), \\
	O_{22}(\tau,\tau_0) &= \cos k(\tau - \tau_0).
\end{align}
In de Sitter spacetime, the scale factor is given by the first line of Eq.~\eqref{scalefactora}. In this case, we find
\begin{align}
	O^{\rm dS}_{11}(\tau,\tau_0) &= \frac{\tau  \cos (k (\tau - \tau_0))}{\tau_0} - \frac{\sin (k (\tau - \tau_0))}{k \tau_0}, \\
	O^{\rm dS}_{12}(\tau,\tau_0) &= \frac{ H ^2 (\tau_0 - \tau) \cos (k (\tau - \tau_0))}{k^2} 
	+ \frac{ H ^2 \left(k^2 \tau \tau_0 + 1\right) \sin (k (\tau - \tau_0))}{k^3}, \\
	O^{\rm dS}_{21}(\tau,\tau_0) &= -\frac{k \sin (k (\tau - \tau_0))}{ H ^2 \tau \tau_0}, \\
	O^{\rm dS}_{22}(\tau,\tau_0) &= \frac{\sin (k (\tau - \tau_0))}{k \tau} + \frac{\tau_0 \cos (k (\tau - \tau_0))}{\tau}.
\end{align}
In radiation dominance, the scale factor is given by the second line of Eq.~\eqref{scalefactora}. With this scale factor, the fundamental matrix is found as
\begin{align}
	O^{\rm RD}_{11}(\tau,\tau_0) &= \frac{\tau_0 \cos\big(k (\tau - \tau_0)\big)}{\tau} + \frac{\sin\big(k (\tau - \tau_0)\big)}{k \tau}, \\
	O^{\rm RD}_{12}(\tau,\tau_0) &= \frac{H^2 \tau_R^4 \sin\big(k (\tau - \tau_0)\big)}{k \tau \tau_0}, \\
	O^{\rm RD}_{21}(\tau,\tau_0) &= \frac{(\tau - \tau_0) \cos\big(k (\tau - \tau_0)\big)}{H^2 \tau_R^4} + \frac{-1 - k^2 \tau \tau_0}{H^2 k \tau_R^4} \sin\big(k (\tau - \tau_0)\big) , \\
	O^{\rm RD}_{22}(\tau,\tau_0) &= \frac{\tau \cos\big(k (\tau - \tau_0)\big)}{\tau_0} - \frac{\sin\big(k (\tau - \tau_0)\big)}{k \tau_0}.
\end{align}

\subsection{Mode Functions for Various Vacuum Choices}

In a Minkowski background, mode functions are uniquely determined up to a phase factor. We obtain
\begin{align}
	u_k(\tau,\tau_0) &= \frac{e^{-ik(\tau-\tau_0)}}{\sqrt{2k}}, \label{def:mdfunc} \\
	v_k(\tau,\tau_0) &= -i \sqrt{\frac{k}{2}} e^{-ik(\tau-\tau_0)}.
\end{align}
During inflation, with the adiabatic vacuum in the remote past, Eq.~\eqref{dstordmd} is found as
\begin{align}
    \lim_{\tau_0 \to -\infty} u^{\rm dS}_k(\tau,\tau_0) &= \frac{i  H  e^{-i k (\tau - \tau_0)} (1 + ik \tau)}{\sqrt{2} k^{3/2}}, \\
    \lim_{\tau_0 \to -\infty} v^{\rm dS}_k(\tau,\tau_0) &= \frac{i \sqrt{k} e^{-i k (\tau - \tau_0)}}{\sqrt{2}  H  \tau}.
\end{align}
During radiation dominance, one may choose a vacuum state at some time after reheating time $\tau_1\geq \tau_R$:
\begin{align}
    u^{\rm RD}_k(\tau,\tau_1) &= 
    \frac{H \tau_R^2 \big(e^{-i q (\tau - \tau_1)} q \tau_1 + \sin\big(q (\tau - \tau_1)\big)\big)}
{\sqrt{2} \, q^{3/2} \, \tau \tau_1}, \label{md_u:rd} \\
    v^{\rm RD}_k(\tau,\tau_1) &= \frac{
q \big(\tau - \tau_1 - i q \tau \tau_1\big) \cos\big(q (\tau - \tau_1)\big) 
- \big(1 + q (-i + q \tau) \tau_1\big) \sin\big(q (\tau - \tau_1)\big)
}{
\sqrt{2} H q^{3/2} \tau_1 \tau_R^2
}. \label{md_v:rd}
\end{align}
In the subhorizon limit $k\tau_R \gg 1$, these mode functions reduce to the Minkowski mode functions rescaled by the scale factor:
\begin{align}
    \lim_{k\tau_R \to \infty} u^{\rm RD}_k(\tau,\tau_{\rm R}) &= \frac{1}{a(\tau)} u_k(\tau), \\
    \lim_{k\tau_R \to \infty} v^{\rm RD}_k(\tau,\tau_{\rm R}) &= a(\tau) v_k(\tau).
\end{align}
This approximation is useful for radiation fields in a thermal state at the initial time of radiation dominance.

\subsection{The mode function Eq.~\eqref{mdinst}}

In the main text, we expanded the graviton field operator with respect to the instantaneous operators.
The mode functions in Eq.~\eqref{mdinst} are found as follows.
First, write
\begin{align}
	Y(\tau_1)= \mathcal O^{\rm RD}(\tau_1,\tau_R) Y(\tau_R),~Y(\tau)= \mathcal O^{\rm RD}(\tau,\tau_R) Y(\tau_R).
\end{align}
Hence,
\begin{align}
	\mathcal O^{\rm RD}(\tau_1,\tau_R)^{-1}Y(\tau_1)= \mathcal O^{\rm RD}(\tau,\tau_R)^{-1}Y(\tau),
\end{align}
and then 
\begin{align}
	Y(\tau_1) = \mathcal O^{\rm RD}(\tau_1,\tau_R) \mathcal O^{\rm RD}(\tau,\tau_R)^{-1}Q(\tau)^{-1} Q(\tau)Y(\tau).
\end{align}
One should note that $ \mathcal O^{\rm RD}(\tau,\tau_1)^{-1} =  \mathcal O^{\rm RD}(\tau_1,\tau)$.
Also,
\begin{align}
	Q(\tau )\hat{Y}(\tau) &=  Q(\tau ) \mathcal{O}_{\rm RD}(\tau,\tau_{\rm R}) \mathcal{O}_{\rm dS}(-\tau_{\rm R},\tau_0)Q(\tau_0)^{-1} Q(\tau_0 )\hat{Y}(\tau_0). \label{insmdt}
\end{align}
Hence, one can read the mode functions and Bogoliubov coefficients as 
\begin{align}
	\begin{pmatrix}
		u_k(\tau_1,\tau) & u^*_k(\tau_1,\tau) \\
		v_k(\tau_1,\tau) & v^*_k(\tau_1,\tau)
	\end{pmatrix}
	= \mathcal O^{\rm RD}(\tau_1,\tau) Q(\tau)^{-1},\label{defmdinsthen}
\end{align}
and 
\begin{align}
	\begin{pmatrix}
		\mu_k(\tau) & \nu_k(\tau) \\
		\nu^*_k(\tau) & \mu^*_k(\tau)
	\end{pmatrix}
	= Q(\tau ) \mathcal{O}_{\rm RD}(\tau,\tau_{\rm R}) \mathcal{O}_{\rm dS}(-\tau_{\rm R},\tau_0)Q(\tau_0)^{-1}.\label{defbt2}
\end{align}

We evaluate Eq.~\eqref{defmdinsthen} and find Eq.~\eqref{mdinst}.
We evaluate Eq.~\eqref{defbt2} and find
\begin{align}
\mu_k &= \frac{1}{4 k^4 \tau \tau_0 \tau_R^2} \Big[
    k \big( i \tau (2 k \tau_0 + i) (-1 + 2 k \tau_R (k \tau_R + i)) \notag \\
    &\quad + \tau_0 (-1 + 2 k \tau_R (k \tau_R + i)) - 2 \tau_R \big) 
    \sin \big( k (-\tau + \tau_0 + 2 \tau_R) \big) \notag \\
    &\quad + \big( 
        -1 + k \big( 
        \tau (2 k \tau_0 + i) (-1 + 2 k \tau_R (k \tau_R + i)) \notag \\
    &\quad + 2 k \tau_R (-i k \tau_0 \tau_R + \tau_0 + \tau_R) + i \tau_0 
    \big) \big) 
    \cos \big( k (-\tau + \tau_0 + 2 \tau_R) \big) \notag \\
    &\quad + k (\tau + \tau_0) \sin \big( k (\tau + \tau_0) \big) 
    + \big( 1 + i k (\tau - \tau_0) \big) 
    \cos \big( k (\tau + \tau_0) \big) \Big], \label{eq:mu_k} \\
\nu_k &= \frac{1}{4 k^4 \tau \tau_0 \tau_R^2} \Big[
    k \big( 
        2 k^2 \tau_R^2 (\tau - \tau_0) 
        + 2 i k \tau_R (\tau + \tau_0) \notag \\
    &\quad - \tau + \tau_0 + 2 \tau_R 
    \big) \sin \big( k (-\tau + \tau_0 + 2 \tau_R) \big) \notag \\
    &\quad + \big( 
        2 k^2 \tau \tau_0 - i k (\tau + \tau_0) - 1 
    \big) \cos \big( k (\tau + \tau_0) \big) \notag \\
    &\quad + \big( 
        -2 i k^3 \tau_R^2 (\tau + \tau_0) 
        + 2 k^2 \tau_R (\tau - \tau_0 - \tau_R) \notag \\
    &\quad + i k (\tau + \tau_0) + 1 
    \big) \cos \big( k (-\tau + \tau_0 + 2 \tau_R) \big) \notag \\
    &\quad + k \big( 
        -\tau_0 + \tau (-1 - 2 i k \tau_0) 
    \big) \sin \big( k (\tau + \tau_0) \big) \Big]. \label{eq:nu_k}
\end{align}
These yield Eqs.~\eqref{nunu} and \eqref{numu} in $\tau_0\to -\infty$ limit.

\section{Summary of Green functions}
\label{appGreen}
The annihilation and creation operators $(\hat b,\hat b^\dagger)$ write $\chi$ as 
\begin{align}
	\hat \chi_{\mathbf p}(\tau_1) = u_p(\tau_1) \hat b_{\mathbf p} + u^*_p(\tau_1) \hat b^\dagger_{- \mathbf p},~ [\hat b_{\mathbf p_1}, \hat b^\dagger_{- \mathbf p_2}] = (2\pi)^3 \delta(\mathbf p_1+\mathbf p_2).
\end{align}
Then Eq.~\eqref{defretG} yields
\begin{align}
	G^R_{p}(\tau_1,\tau_2) = i a^2(\tau_2)(u_p(\tau_1) u^*_p(\tau_2) - u^*_p(\tau_1) u_p(\tau_2) )  \Theta(\tau_1-\tau_2).
\end{align}
The retarded Green function is irrelevant to a choice of a state.
For a given canonical ensemble, the expectation value of the number operator is related to the occupation number $f_{\beta p}$: 
\begin{align}
	{\rm Tr}\left[\hat \varrho  \hat b^{\dagger}_{-\mathbf p_1} \hat b_{\mathbf p_2}  \right] = f_{\beta p_1} (2\pi)^3 \delta(\mathbf p_1 + \mathbf p_2).
\end{align}
Then, Eq.~\eqref{defKelG} yields
\begin{align}
	 G^K_{p}(\tau_1,\tau_2) = a^2(\tau_2)\left[ u_p(\tau_1) u^*_p(\tau_2)+ u^*_p(\tau_1) u_p(\tau_2) \right](1 + 2 f_{\beta p}).
\end{align}
When taking the sub-horizon limit, one can write the positive frequency mode function by
\begin{align}
		u_p(\tau,\tau_0) &= \frac{1}{a(\tau)}\frac{e^{-ip(\tau-\tau_0)}}{\sqrt{2p}}.
\end{align}
Then, we find 
\begin{align}
	G^R_{p}(\tau_1,\tau_2) &= \frac{a(\tau_2)}{a(\tau_1)} \frac{\sin p(\tau_1-\tau_2)}{p}  \Theta(\tau_1-\tau_2),\label{defr}
	\\
	G^K_{p}(\tau_1,\tau_2) &\simeq  \frac{a(\tau_2)}{a(\tau_1)}\frac{\cos p(\tau_1-\tau_2)}{p} (1 + 2 f_{\beta p}).\label{defk}
\end{align}
Here, $f_x = 1/(e^x-1)$ is the Planck distribution and $\Theta$ is the step function. The retarded Green function is independent of the vacuum choice, while the Keldysh Green function depends on it. However, in the sub-horizon limit $p \tau_i \gg 1$, they coincide with the rescaled one in Minkowski spacetime. Note that we assumed that the graviton wavelength $k^{-1}$ is much longer than those of the thermal scalar fields, i.e., $k \beta \ll 1$.

\section{The stimulated emission rate $r_k$}\label{appC}

Eqs.~\eqref{lambdaq} and \eqref{nuq} are written in terms of the mode functions with respect to the instantaneous vacuum~\eqref{mdinst}. 
Integrating these with respect to $\tau_1$ and $\tau_2$, with $\tau_1\sim \tau_2 + \beta/2\pi$, and taking the high temperature limit, $\beta /\tau \to 0$, we eqvaluate Eq.~\eqref{def:r}.
Firstly, we obtain ${\rm Re}[\zeta_k] = \mathcal O(\beta^{-3}),$
and
\begin{align}
\sigma_k = &\frac{e^{-2i k (\tau + \tau_R)} H^2 \pi^2  \tau_R^3}{1800 k^3 \tau^2 \beta^4} \Bigg[
    -i e^{4i k \tau} 
    + i e^{4i k \tau_R} \big(i - 2k\tau\big)^2 \notag \\
    &\quad + e^{2i k (\tau + \tau_R)} \Big(2i + 4k\tau \big(-1 - ik\tau_R\big)\Big) \notag \\
    &\quad - 2 e^{2i k (2\tau + \tau_R)} k\tau_R \Big(
        \operatorname{Ei}(-2i k \tau) - \operatorname{Ei}(-2i k \tau_R)
    \Big) \notag \\
    &\quad - 2 e^{2i k \tau_R} k \big(i - 2k\tau\big)^2 \tau_R \Big(
        \operatorname{Ei}(2i k \tau) - \operatorname{Ei}(2i k \tau_R)
    \Big)
\Bigg],
\end{align}
where Ei is the exponential integral function.
Then, combining these with Eqs.~\eqref{nunu} and \eqref{numu}, we find

	\begin{align}
r_k=&\frac{4}{5} \Bigg[  
2 \Big(1 + x^2 + x_R^4 + (-1 + x_R (x_R - x (2 + x_R^2))) \cos(2x - 2x_R) \Big) \notag \\
& + (-3x + 3x_R + 2x x_R^2 + x_R^3) \sin(2x - 2x_R) \notag \\
& + \operatorname{Ci}(2x_R) \Big( (1 + 2x^2) (-1 + 2x_R^2) \cos(2x_R) + (1 + 2x_R^4)(\cos(2x) + 2x \sin(2x)) \notag \\
& \quad - 2(1 + 2x^2)x_R \sin(2x_R) \Big) \notag \\
& + \operatorname{Ci}(2x) \Big( - (1 + 2x^2) (-1 + 2x_R^2) \cos(2x_R) - (1 + 2x_R^4)(\cos(2x) + 2x \sin(2x)) \notag \\
& \quad + 2(1 + 2x^2)x_R \sin(2x_R) \Big) \notag \\
& + \Big( 2x(1 + 2x_R^4) \cos(2x) - (1 + 2x_R^4) \sin(2x) \notag \\
& \quad - (1 + 2x^2)(2x_R \cos(2x_R) + (-1 + 2x_R^2) \sin(2x_R)) \Big) \big( \operatorname{Si}(2x) - \operatorname{Si}(2x_R) \big)
\Bigg] \notag \\
\Bigg/ & \Big( 5 \Big( 1 + 2x^2 + 2x_R^4 + (-1 + 2x_R(-2x + x_R)) \cos(2x - 2x_R) \notag \\
& \quad + 2(-x + x_R + 2x x_R^2) \sin(2x - 2x_R) \Big) \Big),
\end{align}
where $x_R\equiv k\tau_R$ and $x\equiv k\tau$.
We also find the secular term in the IR limit:
\begin{align}
 \lim_{k\tau\to 0}   r_k = \frac{4}{5} \left[\ln\left( \frac{\tau}{\tau_R} \right) - \frac{2}{3} \left(1 - \frac{\tau_R}{\tau}\right) \left( 2 + \frac{\tau_R^3}{\tau^3} \right) \right]. \label{logfact}
\end{align}
Similar effects were also discussed for the power spectrum of $\hat h$ in Refs.~\cite{Ota:2023iyh,Ota:2022xni,Ota:2022hvh,Chen:2022dah}, whereas $r_k$ in this paper is a correction to the graviton number.

\bibliography{sample.bib}{}
\bibliographystyle{unsrturl}

\end{document}